\documentclass[letterpaper,aps,superscriptaddress,preprint]{revtex4}
\usepackage{graphicx}
\usepackage{amsmath}
\usepackage{epstopdf}
\usepackage{amsfonts}
\usepackage{amssymb}

\begin{document}

\title{Theory of current-driven magnetization dynamics in inhomogeneous ferromagnets}
\author{Yaroslav Tserkovnyak}
\affiliation{Department of Physics and Astronomy, University of California, Los Angeles, California 90095, USA}
\author{Arne Brataas}
\affiliation{Department of Physics, Norwegian University of Science and Technology, NO-7491 Trondheim, Norway}
\author{Gerrit E. W. Bauer}
\affiliation{Kavli Institute of NanoScience, Delft University of Technology, 2628 CJ Delft, The Netherlands}

\begin{abstract}
We give a brief account of recent developments in the theoretical understanding of the interaction between electric currents and inhomogeneous ferromagnetic order parameters. We start by discussing the physical origin of the spin torques responsible for this interaction and construct a phenomenological description. We then consider the electric current-induced ferromagnetic instability and domain-wall motion. Finally, we present a microscopic justification of the phenomenological description of current-driven magnetization dynamics, with particular emphasis on the dissipative terms, the so-called Gilbert damping $\alpha$ and the $\beta$ component of the adiabatic current-driven torque.
\end{abstract}

\date{\today}
\maketitle

\section{Introduction}

Ferromagnetism is a correlated state in which, at sufficiently low temperatures, the electrons align their spins in order to reduce the exchange energy. Below the Curie temperature $T_c$, the free energy $F[\mathbf{M}]$ then attains its minimum at a finite magnetization $M\neq0$ with an arbitrary direction, thus spontaneously breaking the spin-rotational symmetry. Crystal anisotropies that are caused by the spin-orbit interaction and shape anisotropies governed by the magnetostatic dipolar interaction pin the equilibrium magnetization direction to a certain plane or axis. 
At low temperatures, $T\ll T_c$, fluctuations of the magnitude of the magnetization around the saturation value $ M_s$ (the so-called Stoner excitations) become energetically unfavorable. The remaining low-energy long-wavelength excitations are spin waves (or magnons, which, in technical terms, are viewed as Goldstone modes that restore the broken symmetry). These are slowly varying modulations of the magnetization direction in space and time.

A phenomenological description of the slow collective magnetization dynamics without dissipation proceeds from the free energy $F[\mathbf{M}(\mathbf{r})]$ as a functional of the inhomogeneous (and instantaneous) magnetic configuration $\mathbf{M}(\mathbf{r})$ \cite{landauBOOKv9}. The equation of motion\begin{equation}
\frac{\partial\mathbf{M}(\mathbf{r},t)}{\partial t}=\gamma\mathbf{M}(\mathbf{r},t)\times\frac{\delta F[\mathbf{M}]}{\delta\mathbf{M}}
\label{LL}
\end{equation}
preserves the total free energy of the system, since the rate of change of the magnetization is perpendicular to the ``gradient" of the free energy. The functional derivative of the free energy with respect to the local magnetization is called the \textit{effective} field: $\mathbf{H}_{\rm eff}(\mathbf{r},t)=-{\delta F[\mathbf{M}]}/{\delta \mathbf{M}}$. [In the following, we use the abbreviations $\partial _t = \partial / \partial t$ for the partial derivative in time and $\partial _{\mathbf{M}} = {\delta }/{\delta \mathbf{M}}$ for the functional derivative with respect to $\mathbf{M}$.]   In the presence of only an \textit{externally applied} magnetic field 
$\mathbf{H}$, $F[\mathbf{M}]=-\int d^3r\mathbf{M}(\mathbf{r})\cdot\mathbf{H}(\mathbf{r})$, so
$\gamma$ is identified as the effective gyromagnetic ratio. In general, $\mathbf{H}_{\rm eff}$ includes the crystal anisotropy due to spin-orbit interactions, modulation of the exchange energy due to magnetization gradients, and demagnetization fields due to dipole-dipole interactions. The Landau-Lifshitz equation~(\ref{LL}) qualitatively describes many ferromagnetic resonance (FMR) \cite{bhagatPRB74,heinrichAP93} and Brillouin light scattering \cite{camleyJPCM93} experiments. With $\mathbf{M}=M_s\mathbf{m}$, where $\mathbf{m}$ is the magnetic direction unit vector, we may rewrite Eq.~(\ref{LL}) as
\begin{equation}
\partial_t\mathbf{m}(\mathbf{r},t)=-\gamma\mathbf{m}(\mathbf{r},t)\times\mathbf{H}_{\rm eff}(\mathbf{r},t)\,.
\label{ll}
\end{equation}
By definition, the effective field on the right-hand side of Eq.~(\ref{ll}) is determined by the instantaneous magnetic configuration. This can be true only if the motion is so slow that all relevant microscopic degrees of freedom manage to immediately readjust themselves to the varying magnetization. If this is not the case, the effective field acquires a finite time lag that to lowest order in frequency can be schematically expanded as
\begin{equation}
\tilde{\mathbf{H}}_{\rm eff}\to-\partial_\mathbf{M}F[\mathbf{M}(\mathbf{r},t-\tau)]\approx\mathbf{H}_{\rm eff}-\tau(\partial_t\mathbf{M}\cdot\partial_\mathbf{M})\mathbf{H}_{\rm eff}\,,
\end{equation}
where $\tau$ is a characteristic delay time. This dynamic correction to the instantaneous effective field, $\delta\mathbf{H}_{\rm eff}$, leads to a new term in the equation of motion $\propto\mathbf{m}\times\delta\mathbf{H}_{\rm eff}$. Although in general nonlocal and anisotropic \cite{kunesPRB02}, it makes sense to identify first the simplest, i.e., local and isotropic contribution. We can then construct two new terms out of the vectors $\mathbf{m}$ and $\partial_t\mathbf{m}$. The first one, $\propto\mathbf{m} \times \partial_t\mathbf{m}$, is dissipative, meaning that it is odd under time reversal (i.e., under the transformation $t\to-t$, $\mathbf{H}\to-\mathbf{H}$, and $\mathbf{m}\to-\mathbf{m}$), and thus violates the time-reversal symmetry of the Landau-Lifshitz equation (\ref{ll}). This argument leads to the Landau-Lifshitz-Gilbert (LLG) equation \cite{gilbertPR55,gilbertIEEEM04}:
\begin{equation}
\partial_t\mathbf{m}(\mathbf{r},t)=-\gamma\mathbf{m}(\mathbf{r},t)\times\mathbf{H}_{\rm eff}(\mathbf{r},t)+\alpha\mathbf{m}(\mathbf{r},t)\times\partial_t\mathbf{m}(\mathbf{r},t)\,,
\label{llg}
\end{equation}
introducing the phenomenological  Gilbert damping constant $\alpha$. The second local and isotropic term linear in $\partial_t\mathbf{m}$ and perpendicular to $\mathbf{m}$, that can be composed out of $\mathbf{m}$ and $\partial_t\mathbf{m}$ is proportional to $\partial_t\mathbf{m}$ and can be combined with the left-hand side of the LLG equation. In principle, any physical process that contributes to the Gilbert damping can thus also renormalize the gyromagnetic ratio $\gamma$. The latter should therefore be interpreted as an effective parameter in the equation of motion (\ref{llg}). The second law of thermodynamics requires that $\alpha\gamma\geq0$, which guarantees that the dissipation of energy $P\propto\mathbf{H}_{\rm eff}\cdot\partial_t\mathbf{m}\geq0$ (assuming the magnetization dynamics are slow and isolated from any external sinks of entropy). Since the implications of small modifications of the gyromagnetic ratio are minor, we will be mainly concerned with the Gilbert damping constant $\alpha$. Furthermore, we will focus on  dissipative effects due to spin dephasing by magnetic or spin-orbit impurities \cite{heinrichPSS67,sinovaPRB04,tserkovAPL04,koopmansPRL05,hankiewiczPRB07,skadsemPRB07}, noting that many other Gilbert damping mechanisms have been proposed in the past \cite{amentPR55,korenmanPRB72,lutovinovJETP79,solontsovPLA93,suhlIEEEM98,dobinPRL03,steiaufPRB05,gilmorePRL07}. According to the fluctuation-dissipation theorem, the dissipation, whatever its microscopic origin is, must be accompanied by a stochastic contribution $\mathbf{h}(\mathbf{r},t)$ to the effective field.  Assuming Gaussian statistics with a white noise correlator \cite{brownPR63}, which is valid in the classical limit with characteristic frequencies that are sufficiently small compared to thermal energies:
\begin{equation}
\langle h_i(\mathbf{r},t)h_j(\mathbf{r}^\prime,t^\prime)\rangle=2k_BT\frac{\alpha}{\gamma M_s}\delta_{ij}\delta(\mathbf{r}-\mathbf{r}^\prime)\delta(t-t^\prime)\,.
\label{hh}
\end{equation}
Eqs.~(\ref{LL})-(\ref{hh}) form the standard phenomenological basis for understanding dynamics of ferromagnets \cite{bhagatPRB74,heinrichAP93,camleyJPCM93}, in the absence of an applied current or voltage bias.

\section{Current-driven magnetization dynamics}

In order to understand recent experiments on current-biased magnetic multilayers \cite{tsoiPRL98,myersSCI99,sunJMMM99,wegroweEPL99,katinePRL00,tsoiNAT00,grollierAPL01,kiselevNAT03,krivorotovSCI05} and nanowires \cite{yamaguchiPRL04,yamanouchiPRL06,beachPRL06,thomasNAT06,hayashiPRL06,meierPRL07,hayashiNATP07,yamanouchiSCI07}, Eq.~(\ref{llg}) has to be modified \cite{bergerJAP78,bergerPRB96,slonczewskiJMMM96}. The leading correction has to take into account the finite divergence of the spin-current density in conducting ferromagnets, with magnetization texture that has to be brought into compliance with the conservation of angular momentum. We have to introduce a new term
\begin{equation}
s_0\,\partial_tm_i|_{\rm torque}=\boldsymbol{\nabla}\cdot\mathbf{j}_i
\end{equation}
in the presence of a current density $\mathbf{j}_i$ for spin-$i$ component, where $s_0$ is the total equilibrium spin density along $-\mathbf{m}$ (the minus sign takes into account that electron spin and magnetic moment point in opposite directions). By adding this term to the right-hand side of Eq.~(\ref{llg}) as a contribution to $\partial_tm_i$, we assume that the angular momentum lost in the spin current is fully added to the magnetization. This is called the spin-transfer torque \cite{slonczewskiJMMM96,bazaliyPRB98,capellePRL01,liPRL04}. The simplest approximation for the spin-current density in the bulk of isotropic ferromagnets \cite{bazaliyPRB98,fernandezPRB04,liPRL04} is $\mathbf{j}_i=\mathcal{P}\mathbf{j}m_i$, where $\mathcal{P}$ is a material-dependent constant that converts charge-current density $\mathbf{j}$ into spin-current density. The underlying assumption here is that spins are carried by the electric current such that the spin-polarization axis adiabatically follows the local magnetization direction. This is the case for  a large exchange field that varies slowly in space. This condition is fulfilled very well in transition-metal ferromagnets. The spin conversion factor
\begin{equation}
\mathcal{P}=\frac{\hbar}{2e}\frac{\sigma_\uparrow-\sigma_\downarrow}{\sigma_\uparrow+\sigma_\downarrow}
\label{P}
\end{equation}
characterizes the polarization of the spin-dependent conductivity $\sigma_s$ ($s=\uparrow$ or $s=\downarrow$) with $\uparrow$ chosen along $-\mathbf{m}$. Hence
\begin{equation}
\partial_t\mathbf{m}=-\gamma\mathbf{m}\times\mathbf{H}_{\rm eff}+\frac{\mathcal{P}}{s_0}(\mathbf{j}\cdot\boldsymbol{\nabla})\mathbf{m}\,,
\label{llgs}
\end{equation}
where we took into account local charge neutrality by $\boldsymbol{\nabla}\cdot\mathbf{j}=0$. 

The phenomenological Eq.~(\ref{llgs}) is ``derived" without taking into account spin relaxation processes. Its inclusion requires some care since both Gilbert damping and spin-transfer torque are nontrivially affected \cite{zhangPRL04,xiaoPRB06,tserkovPRB06md,kohnoJPSJ06,duinePRB07,bergerPRB07}. Spin relaxation is generated by impurities with potentials that do not commute with the spin density operator, such as a quenched random magnetic field or spin-orbit interaction associated with randomly distributed non-magnetic impurities \cite{zhangPRL04,tserkovPRB06md,skadsemPRB07,kohnoJPSJ06,duinePRB07}. In the absence of an applied current $\mathbf{j}$, imperfections with potentials that mix the spin channels contribute to the Gilbert constant $\alpha$ \cite{heinrichPSS67,sinovaPRB04,tserkovAPL04,skadsemPRB07,kohnoJPSJ06}. It is instructive to interpret the right-hand side of the equation of motion (\ref{llgs}) as an analytic expansion of driving and damping torques in $\boldsymbol{\nabla}$ and $\partial_t$. The LLG equation (\ref{llg}) corresponds to the most general (local and isotropic) expression for the damping to the zeroth order in $\boldsymbol{\nabla}$ and first order in $\partial_t$. We will not be concerned with higher order terms in $\partial_t$, since the characteristic frequencies of magnetization dynamics are typically small on the scale of the relevant microscopic energies, at least in metallic systems. The contribution to the effective field due to a finite magnetic stiffness \cite{landauBOOKv9} is proportional to $\boldsymbol{\nabla}^2$. In the presence of inversion symmetry, the terms proportional to $\boldsymbol{\nabla}$ cannot appear without applied electric currents. In the following, we focus on the current-driven terms linear in $\boldsymbol{\nabla}$, assuming that the spatial variations in the magnetization direction are sufficiently smooth to rule out higher-order contributions. The dynamics of isotropic spin-rotationally invariant ferromagnets can then in general be described by the phenomenological equation of motion \cite{zhangPRL04,thiavilleEPL05,tserkovPRB06md}
\begin{equation}
\partial_t\mathbf{m}=-\gamma\mathbf{m}\times\mathbf{H}_{\rm eff}+\alpha\mathbf{m}\times\partial_t\mathbf{m}+\frac{\mathcal{P}}{s_0}\left(1-\beta\mathbf{m}\times\right)(\mathbf{j}\cdot\boldsymbol{\nabla})\mathbf{m}\,,
\label{llab}
\end{equation}
in which $\alpha$ and $\beta$ characterize those terms that break time-reversal symmetry. Both arise naturally in the presence of spin-dependent impurities \cite{zhangPRL04,tserkovPRB06md,kohnoJPSJ06}. Even though in practice $\beta\ll1$, it gives an important correction to the current-driven spin-transfer torque \cite{zhangPRL04,thiavilleEPL05,tserkovPRB06md}, as discussed in more detail below. For the special case of $\alpha=\beta$: Eq.~(\ref{llab}) can then be rewritten (after multiplying it by $1+\alpha\mathbf{m}\times$ on the left) as
\begin{equation}
\partial_t\mathbf{m}=-\gamma^\ast\mathbf{m}\times\mathbf{H}_{\rm eff}+\alpha\gamma^\ast\mathbf{m}\times\mathbf{H}_{\rm eff}\times\mathbf{m}+\frac{\mathcal{P}}{s_0}(\mathbf{j}\cdot\boldsymbol{\nabla})\mathbf{m}\,,
\label{lla}
\end{equation}
where $\gamma^\ast=\gamma/(1+\alpha^2)$. The dissipative term proportional to $\mathbf{m}\times\mathbf{H}_{\rm eff}\times\mathbf{m}$ is called the Landau-Lifshitz damping. Eq. (\ref{llab}) cannot be transformed into equation (\ref{lla}) if $\alpha\neq\beta$ [in which case Eq.~(\ref{lla}) necessarily retains a $\beta$ term]. A special feature of Eq.~(\ref{lla}) appears when $\mathbf{H}_{\rm eff}(\mathbf{m})$ is time independent and translationally invariant. A general solution $\mathbf{m}(\mathbf{r},t)$ in the absence of an electric current  $\mathbf{j}=0$ (such as a static domain wall or a spin wave) can be used to construct a solution
\begin{equation}
\tilde{\mathbf{m}}(\mathbf{r},t)=\mathbf{m}(\mathbf{r}+\mathcal{P}\mathbf{j}t/s_0,t)
\label{gi}
\end{equation}
of Eq.~(\ref{lla}) for an arbitrary uniform $\mathbf{j}$. This unique feature of the solutions of Eq.~(\ref{llab}) only arises for $\alpha=\beta$.

Interestingly, the argument above has been turned around by Barnes and Maekawa \cite{barnesPRL05}, who find that Galilean invariance of a system would dictate $\alpha=\beta$. Galilean invariance requires the existence of solutions of the form $\tilde{\mathbf{m}}(\mathbf{r}-\mathbf{v}t)$, where $\tilde{\mathbf{m}}(\mathbf{r})$ is an arbitrary static solution (say, a domain wall) and $\mathbf{v}$ is an arbitrary velocity.  As explained above, this is only possible when $\alpha=\beta$. However, the general validity of the Galilean invariance assumption for the current-carrying state needs to be discussed in more detail from a microscopic point of view. The Galilean invariance argument \cite{barnesPRL05} implies that the bias-induced electron drift exactly corresponds to the domain-wall velocity, since otherwise electron motion would persist in the frame that moves with the domain wall. Referring to Eq.~(\ref{gi}), we must, therefore, identify $\mathbf{v}=-\mathcal{P}\mathbf{j}/s_0$ with the average electron drift velocity in the presence of the current $\mathbf{j}$. We argue in the following that this is indeed true in certain special limits, but is not generic, however. In the itinerant Stoner model for ferromagnets, the spin-dependent Drude conductivity reads $\sigma_s\propto n_s\tau_s$, where $n_s$ and $\tau_s$ are the densities and scattering times of spin $s$, respectively. When there is no asymmetry between the scattering times, $\tau_\uparrow=\tau_\downarrow$, $\mathbf{v}$ indeed equals the electron drift velocity and Galilean invariance is effectively fulfilled. In general, however, since the spin dependence of wave functions and densities of states for the electrons at the Fermi energy lead to different scattering cross sections in conducting ferromagnets, the equality between $-\mathcal{P}\mathbf{j}/s_0$ and the average drift velocity disappears. In the simplest model of perturbative white-noise impurity potentials, for example,  $1/\tau_s\propto\nu_s$, where $\nu_s$ is the spin-dependent density of states. Assuming parabolic free-electron bands and  weak ferromagnets, in which the ferromagnetic exchange splitting is much less than the Fermi energy, the domain-wall velocity $\mathbf{v}=-\mathcal{P}\mathbf{j}/s_0$ actually becomes $2/3$ of the average drift velocity $\bar{\mathbf{v}}$, 
\begin{equation}
\mathbf{v}=-\frac{\mathcal{P}\mathbf{j}}{s_0}=\frac{n_\uparrow\tau_\uparrow-n_\downarrow\tau_\downarrow}{n_\uparrow\tau_\uparrow+n_\downarrow\tau_\downarrow}\frac{n_\uparrow\mathbf{v}_\uparrow+n_\downarrow\mathbf{v}_\downarrow}{n_\uparrow-n_\downarrow}\approx\frac{n_\uparrow/\nu_\uparrow-n_\downarrow/\nu_\downarrow}{(n_\uparrow-n_\downarrow)(\nu_\uparrow+\nu_\downarrow)/2}\bar{\mathbf{v}}\approx\frac{2}{3}\bar{\mathbf{v}}\,.
\end{equation}
 Here, $\mathbf{v}_s$ is the spin-$s$ electron drift velocity, and we used the relation $\nu_s\propto n_s^{1/3}$, which is valid in three dimensions. Clearly, the potential disorder breaks Galilean invariance. An identity of $\alpha$ and $\beta$ can therefore not be deduced from general symmetry principles. Furthermore,  spin-orbit interaction or magnetic disorder that strongly affect the values of $\alpha$ and $\beta$ (see below) also break Galilean invariance at the level of the microscopic Hamiltonian. Nevertheless, for itinerant ferromagnets we show below that $\alpha\sim\beta$  (where by $\sim$ we mean ``of the order"), with $\alpha\approx\beta$ in the simplest model of weak and isotropic spin-dephasing impurities \cite{tserkovPRB06md}, which implies that deviations from translational invariance are not very important in metallic ferromagnets, such as transition metals and their alloys, in which the Stoner model is applicable. Very recently, two independent groups measured $\alpha\approx\beta$ in permalloy nanowires \cite{hayashiPRL06,meierPRL07}. 

Let us also consider the $s-d$ model of ferromagnetism. When, as is usually done, the $d$-orbital lattice is assumed spatially locked, Galilean invariance is broken even in the absence of disorder. In this case, the ratio $\alpha/\beta$ deviates strongly from unity, although it remains to be relatively insensitive to the strength of spin-dependent impurities. In other words, $\alpha$ and $\beta$ scale similarly with the strength of spin-dephasing processes, and their ratio appears to be determined mainly by band-structure effects and the nature (rather than the strength) of the disorder \cite{tserkovPRB06md,kohnoJPSJ06}. A predictive material-dependent theory of magnetization damping and current-induced domain wall motion that transcends the toy models mentioned above is beyond the scope of our paper. 

The form of the equation of motion (\ref{lla}) for the special case $\alpha=\beta$ has also triggered the suggestion \cite{stilesPRB07} that the Landau-Lifshitz form of damping, $\propto\mathbf{m}\times\mathbf{H}_{\rm eff}\times\mathbf{m}$, is more natural than the Gilbert form, $\propto\mathbf{m}\times\partial_t\mathbf{m}$. In our opinion, however, such a distinction is purely semantic. Both forms are odd under time reversal, and one can easily imagine simple models in which either form arises more naturally than the other: For example, a Bloch-like $T_2$ relaxation added to the Stoner model naturally leads to a Landau-Lifshitz form of damping \cite{tserkovPRB06md}, whereas the dynamic interface spin pumping \cite{tserkovPRL02sp,tserkovRMP05} very generally obeys the Gilbert damping form. Moreover, mathematically, both equations are identical (in the absence of any additional torques), since we have shown above that the Landau-Lifshitz form of damping follows from the Gilbert one simply by multiplying both sides of the LLG form by $1+\alpha\mathbf{m}\times$ from the left (and vice versa by $1-\alpha\mathbf{m}\times$). Only at the special point $\alpha=\beta$, the Landau-Lifshitz form (\ref{lla}) does not involve the ``$\beta$ term." In that limit, it may be a more transparent expression for the equation of motion. On the other hand, we noted above that in general $\alpha\neq\beta$ and the ratio $\alpha/\beta$ depends on material and sample. The current-driven dynamics of domain walls and other spatially nonuniform magnetization distributions turn out to be very sensitive to small deviations of $\alpha/\beta$ from unity, which strongly reduces any advantage a Landau-Lifshitz damping formulation might have over the Gilbert phenomenology. In general, we therefore prefer to use the Gilbert phenomenology.

Under time reversal, the electric current as well as the magnetization vector change sign and the adiabatic current-induced torque is symmetric, thus nondissipative. The Ohmic dissipation generated by this current does not depend on the magnetization texture in this limit and is intentionally disregarded. Saslow \cite{saslowCM07} prefers to discuss a torque driven by voltage rather current, which, after inserting Ohm's law for the current, becomes odd under time reversal and thus appears dissipative (see Ref.~\cite{smithCM07} for another discussion of this point). Obviously, the $\beta$ correction torque is odd for the current-biased and even for the voltage-biased configurations. The current-bias picture appears to be more natural, since it reflects the absence of additional dissipation by the magnetization texture in the adiabatic limit as well as the close relation between the $\beta$ correction and the Gilbert dissipation.

\section{Current-driven instability of ferromagnetism}

Let us now pursue some special aspects of the solutions of the phenomenological Eq.~(\ref{llab}), highlighting the role of various parameters, before we discuss the microscopic derivation of the magnetization dynamics in Sec.~\ref{mt}. It is interesting, for example, to investigate the possibility to destabilize a single-domain ferromagnet by sufficiently large spin torques \cite{bazaliyPRB98,fernandezPRB04}. We consider a homogeneous ferromagnet with an easy-axis anisotropy along the $x$ axis, characterized by the anisotropy constant $K$, and an easy-plane anisotropy in the $xy$ plane, with the anisotropy constant $K_\perp$, see Fig.~\ref{neel}. Typically, the anisotropies originate from the demagnetization fields: For a ferromagnetic wire, for example, the magnetostatic energy is lowest when the magnetization is in the wire direction, so that there are no stray field lines outside the ferromagnet. The effective field governing magnetization dynamics is then given by
\begin{equation}
\mathbf{H}_{\rm eff}=(H+Km_x)\mathbf{x}-K_\perp m_z\mathbf{z}+A\nabla^2\mathbf{m}\,,
\label{H}
\end{equation}
where we also included an applied field $H$ along the $x$ axis and the exchange coupling parametrized by the stiffness constant $A$. $A,H,K,K_\perp\geq0$. We then look for spin-wave solutions of the form
\begin{equation}
\mathbf{m}=\mathbf{x}+\mathbf{u}e^{i(\mathbf{q}\cdot\mathbf{r}-\omega t)}\,,
\end{equation}
plugging it into Eq.~(\ref{llab}) with the effective field (\ref{H}) in the presence of a constant current density $\mathbf{j}$, and linearizing it with respect to small deviations $\mathbf{u}$. When $\alpha=0$ and $\mathbf{j}=0$, we recover the usual spin-wave dispersion (Kittel formula):
\begin{equation}
\omega_0(q)=\gamma\sqrt{(H+K+Aq^2)(H+K+K_\perp+Aq^2)}\,.
\end{equation}
A finite $\alpha\gamma>0$ results in a negative ${\rm Im}\,\omega(q)$, as required by the stability of the ferromagnetic state. A sufficiently large electric current may, however, reverse the sign of ${\rm Im}\,\omega(q)$ for certain wave vectors $q$, signaling the onset of an instability. The critical value of the current for the instability corresponds to the condition ${\rm Im}\,\omega(q)=0$. Straightforward manipulations based on Eqs.~(\ref{llab}) and (\ref{H}) show that this condition is satisfied when
\begin{equation}
\frac{\mathcal{P}}{s_0}\left(1-\frac{\beta}{\alpha}\right)(\mathbf{q}\cdot\mathbf{j}_c)=\pm\omega_0(q)\,.
\end{equation}
which leads to a critical current density 
\begin{equation}
j_c=\frac{j_{c0}}{|1-\beta/\alpha|}\,.
\label{jc}
\end{equation}
$j_{c0}$ is the lowest current satisfying equation $(\mathcal{P}/s_0)(\mathbf{q}\cdot\mathbf{j}_{c0})=\omega_0(q)$ for some $\mathbf{q}$, where the left-hand side can be loosely interpreted as the current-induced Doppler shift to the natural frequency given by the right-hand side \cite{fernandezPRB04}. According to Eq.~(\ref{jc}), a current-driven instability is absent when $\alpha=\beta$. This conclusion is in line with the arguments leading to Eq.~(\ref{gi}): For the special case of $\alpha=\beta$, a spin-wave solution in the presence of a finite current density $\mathbf{j}$ would acquire a frequency boost proportional to $\mathbf{q}\cdot\mathbf{j}$, but with a stable amplitude. Note that in general the onset of the current-driven ferromagnetic instability is significantly modified by the existence of $\beta$ even with $\beta\ll1$, provided that the ratio $\beta/\alpha$ is appreciable. In fact, $\alpha$ is typically measured to be $\sim0.001-0.01$, and the existing microscopic theories \cite{tserkovPRB06md,kohnoJPSJ06} predict $\beta$ to be not too different from $\alpha$.

\section{Current-driven domain-wall motion}

Even more interesting phenomena are associated with the effect of the applied electric current on a stationary domain-wall. In particular, we wish to discuss how the spin torques move and distort a domain wall. These questions date back about three decades \cite{bergerJAP78}, although only relatively recently they sparked an intense activity by several groups \cite{tataraPRL04,barnesPRL05,thiavilleEPL05,tserkovPRB06md,kohnoJPSJ06,duinePRB07}. This is motivated by the growing number of intriguing experiments \cite{yamaguchiPRL04,yamanouchiPRL06,beachPRL06,thomasNAT06,hayashiPRL06,hayashiNATP07,yamanouchiSCI07} as well as the promise of practical potential, such as in the so-called racetrack memory \cite{parkinUSP06} or magnetic logics \cite{allwoodSCI05}. Current-induced domain-wall motion is a central topic of the present review.

\begin{figure}[tbp]
\includegraphics[width=0.8\linewidth,clip=]{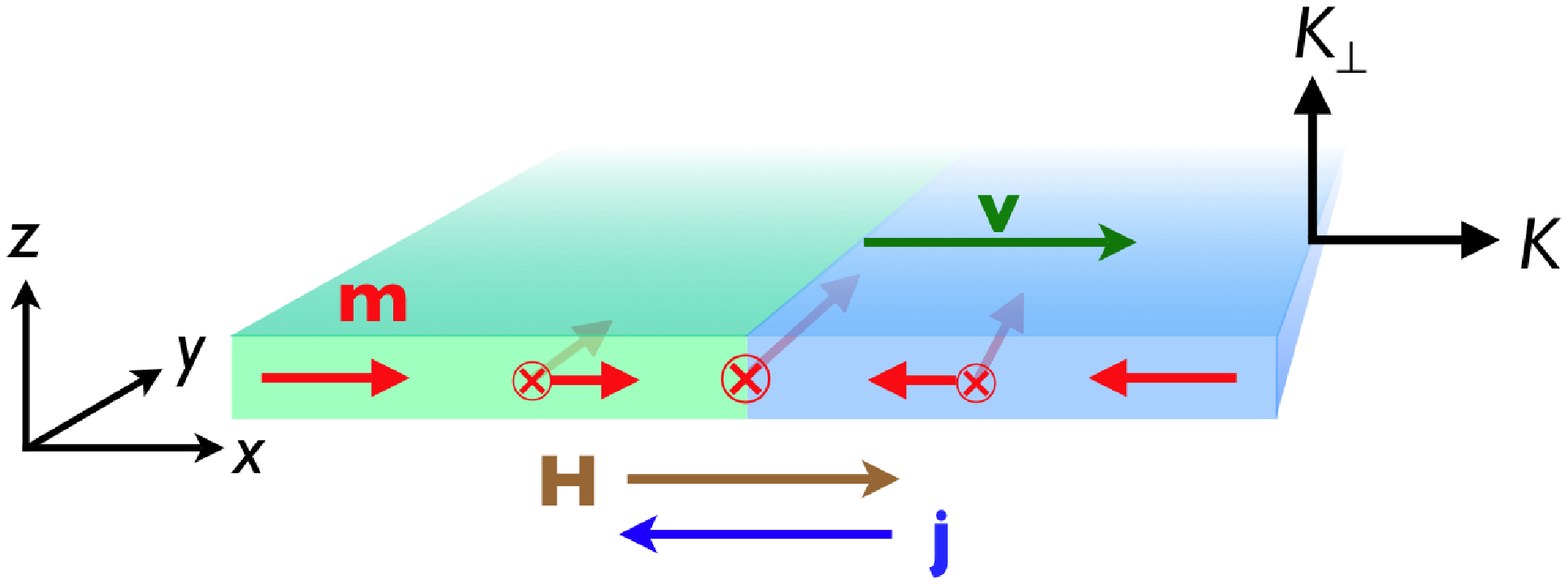}
\caption{Transverse head-to-head (N{\'e}el) domain wall parallel to the $y$ axis in the easy $xy$ plane. The uniform magnetization has two stable solutions $\mathbf{m}=\pm\mathbf{x}$ along the easy axis $x$, which is characterized by the anisotropy constant $K$. These are approached far away from the domain wall: $\mathbf{m}\to\pm\mathbf{x}$ at $x=\mp\infty$, respectively. In equilibrium, the magnetization direction $\mathbf{m}$ is forced into the $xy$ plane by the easy-plane anisotropy parametrized by $K_\perp$. A weak magnetic field $\mathbf{H}$ or electric current $\mathbf{j}$ applied along the $x$ axis can induce a slow domain-wall drift along the $x$ axis, during which the magnetization close to the domain wall is tilted slightly out of the $xy$ plane. At larger $\mathbf{H}$ or $\mathbf{j}$ (above the so-called Walker threshold), the magnetization is significantly  pushed out of the $xy$ plane and undergoes precessional motion during the drift. In the moving frame, the magnetization profile may remain still close to the equilibrium one.}
\label{neel}
\end{figure}

For not too strong driving currents and in the absence of any significant transverse dynamics, one can make progress analytically by using the one-dimensional Walker ansatz, which was first employed in studies of magnetic-field driven domain-wall dynamics \cite{schryerJAP74}. This approach has proven useful in the present context as well \cite{liPRB04dw,thiavilleJAP04,tataraPRL04}. The key idea is to approximately capture the potentially complex domain-wall motion by few parameters describing the displacement of its center and a net distortion of the domain-wall structure. In a quasi-one-dimensional set-up, such as a narrow magnetic wire, the domain wall is constrained to move along a certain axis, whereas the transverse dynamics are suppressed. This regime is relevant for a number of existing experiments, although it should be pointed out that the common vortex-type domain walls do not necessarily fall into this category. Let us consider an idealized situation with an effective field (\ref{H}) and an equilibrium domain wall magnetization in the $xy$ plane. The magnetization prefers to be collinear with the $x$ axis due to the easy-axis anisotropy $K$. A transverse head-to-head domain wall parallel to the $y$ axis corresponds to a magnetization direction that smoothly rotates in the $xy$ plane between $\mathbf{x}$ at $x\to-\infty$ and $-\mathbf{x}$ at $x\to\infty$, as sketched in Fig.~\ref{neel}.

The collective domain-wall dynamics can be described by the center position $X(t)$ and an out-of-plane tilting angle $\Phi(t)$. [For a more technically-interested reader, we note that in the effective treatment of Ref.~\cite{tataraPRL04}, these variables are canonically conjugate.] There is also a width distortion, but that is usually considered less important. When $H<K$, the two uniform stable states are $\mathbf{m}=\pm\mathbf{x}$. When $H=0$, a static transverse head-to-head domain-wall solution centered at $x=0$ is given by
\begin{equation}
\varphi(x)\equiv0\,,\,\,\,\ln\tan\frac{\theta(x)}{2}=\frac{x}{W}\,,
\label{dws}
\end{equation}
where position-dependent angles $\varphi$ and $\theta$ parametrize the magnetic configuration:
\begin{equation}
\mathbf{m}=(m_x,m_y,m_z)=(\cos\theta,\sin\theta\cos\varphi,\sin\theta\sin\varphi)\,.
\end{equation}
$W=\sqrt{A/K}$ is the wall width, which is governed by the interplay between the stiffness $A$ that tends to smooth the wall extent and the easy-axis anisotropy $K$ that tends to sharpen the wall.

The external magnetic field $H$ or the current density $j$ along the $x$ axis disturb the static solution (\ref{dws}), distorting the domain-wall structure and displacing its position. At weak field and current biases, magnetic dynamics can be captured by the Walker ansatz \cite{schryerJAP74,liPRB04dw}:
\begin{equation}
\varphi(\mathbf{r},t)\equiv\Phi(t)\,,\,\,\,\ln\tan\frac{\theta(\mathbf{r},t)}{2}\equiv\frac{x-X(t)}{\tilde{W}(t)}\,.
\label{WA}
\end{equation}
Here, it is assumed that the driving perturbations ($H$ and $j$) are not too strong, such that the wall preserves its shape, except for a small change of its width $\tilde{W}(t)$ and a uniform out-of-plane tilt angle $\Phi(t)$. $X(t)$ parametrizes the net displacement of the wall along the $x$ axis. Note that although $\varphi$ is assumed to be spatially uniform, it has an effect on the magnetization direction only when $\mathbf{m}\neq\pm\mathbf{x}$, i.e., only near the wall center. A more detailed discussion concerning the range of validity of this approximation can be found in Ref.~\cite{schryerJAP74}. Inserting the ansatz (\ref{WA}) into the equation of motion (\ref{llab}) with $\mathbf{j}=j\mathbf{x}$ (since the other current directions do not couple to the wall), and using Eq.~(\ref{H}) for the effective field, one finds \cite{liPRB04dw,thiavilleEPL05}
\begin{align}
\dot{\Phi}+\frac{\alpha \dot{X}}{\tilde{W}}&=\gamma H-\frac{\beta\mathcal{P}j}{s_0\tilde{W}}\,,\nonumber\\
\frac{\dot{X}}{\tilde{W}}-\alpha\dot{\Phi}&=\frac{\gamma K_\perp\sin2\Phi}{2}-\frac{\mathcal{P}j}{s_0\tilde{W}}\,,\nonumber\\
\tilde{W}&=\sqrt{\frac{A}{K+K_\perp\sin^2\Phi}}\,.
\label{was}
\end{align}
It is easy to verify that the static solution (\ref{dws}) is consistent with these equations when $H=0$ and $j=0$. Two different dynamic regimes can be distinguished based on Eqs.~(\ref{was}): When the driving forces are weak, a slightly distorted wall moves at a constant speed, $\dot{X}={\rm const}$, and constant tilt angle, $\dot{\Phi}=0$ (assuming constant $H$ and $j$). The corresponding Walker ansatz (\ref{WA}) then actually provides the exact solution, which is approached at long times after the constant driving field and/or current are switched on \cite{schryerJAP74,liPRB04dw}. Beyond certain critical values of $H$ or $j$, called Walker thresholds, however, no solution with constant angle $\Phi$ and constant velocity $\dot{X}$ exist. Both undergo periodic oscillations in time, albeit with a finite average drift velocity $\langle\dot{X}\rangle\neq0$. In the spacial case of $\alpha=\beta$, Eqs.~(\ref{was}) are exact at arbitrary dc currents when $H=0$: According to Eq.~(\ref{gi}), the static domain-wall solution (of an arbitrary domain-wall shape) then simply moves with velocity $-\mathcal{P}j/s_0$ without any distortions. When $\beta\neq\alpha$, the Walker threshold current diverges when $\beta$ approaches $\alpha$, reminiscent of the critical current (\ref{jc}) discussed in the previous section.

For subthreshold fields and currents with $\Phi(t)\to{\rm const}$ as $t\to\infty$, the steady state terminal velocity is given by \cite{zhangPRL04}
\begin{equation}
v=\dot{X}(t\to\infty)=\frac{\gamma H\tilde{W}-\beta\mathcal{P}j/s_0}{\alpha}\,.
\label{v}
\end{equation}
In particular, when $j=0$, the wall depicted in Fig.~\ref{neel} moves along the direction of the applied magnetic field $H$ in order to decrease the free energy \cite{schryerJAP74}. Let us in the following focus on the current-driven dynamics with $H=0$. At a finite but small $j$, the wall is slightly compressed according to 
\begin{equation}
1-\frac{\tilde{W}}{W}\approx\frac{(\mathcal{P}j/s_0)^2}{2\gamma^2AK_\perp}\left(1-\frac{\beta}{\alpha}\right)^2\,,
\end{equation}
where $W=\sqrt{A/K}$ is the equilibrium width. When $\alpha=\beta$, the domain-wall velocity $v\to-\mathcal{P}j/s_0$. In this case, if we consider the electron spins following the magnetization direction from $\pm\mathbf{m}$ to $\mp\mathbf{m}$ on traversing the domain wall with current density $j$, the entire angular momentum change is transferred to the domain-wall displacement. In this sense, the ratio $\beta/\alpha$ can be loosely interpreted as a spin-transfer efficiency from the current density to the domain-wall motion.

\begin{figure}[tbp]
\includegraphics[width=0.8\linewidth,clip=]{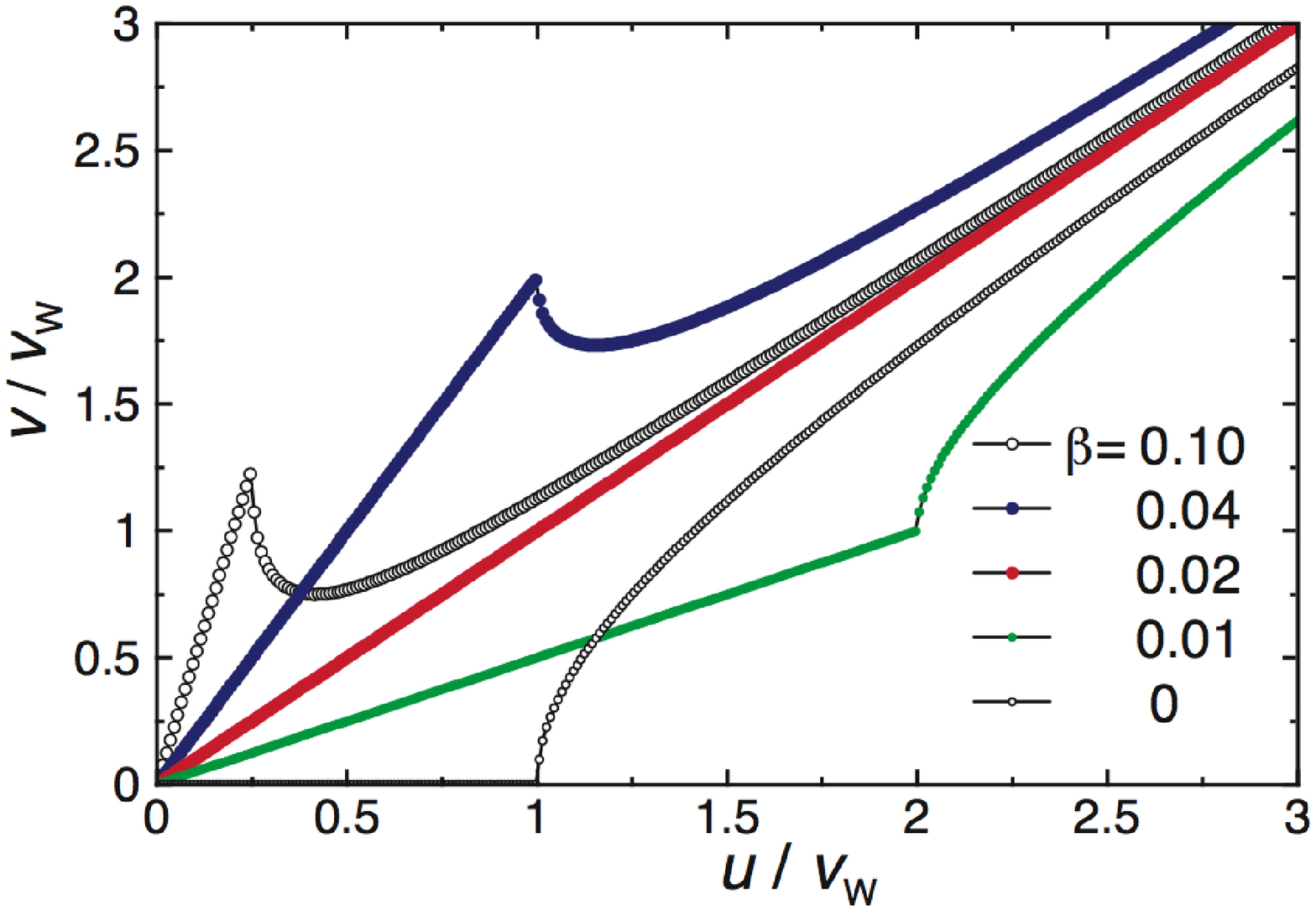}
\caption{Average current-driven domain-wall velocity $v$ numerically calculated using the Walker ansatz [Eqs.~(\ref{was})] in Ref.~\cite{thiavilleEPL05}. Here, the domain-wall width has been approximated by its equilibrium value, $\tilde{W}\approx W$, assuming $K_\perp\ll K$. The curves are  very similar to the full micromagnetic simulations~\cite{thiavilleEPL05}. $u=-\mathcal{P}j/s_0$ has the units of velocity (proportional to electron drift velocity) and $v_w=\gamma K_\perp \zeta/2$ is its value for $j=j_{t0}$. The length $\zeta\approx W$, if we assume $K_\perp\ll K$ (as was done in this calculation), while $\zeta\approx\sqrt{2A/K_\perp}$ in the opposite limit, $K_\perp\gg K$, which is relevant for a thin-film with large demagnetization anisotropy $K_\perp=4\pi M_s$ (in which case $\zeta$ is called exchange length) \cite{liPRB04dw}. $\alpha=0.02$, and we refer to Ref.~\cite{thiavilleEPL05} for the remaining details.}
\label{num}
\end{figure}

Only when $ \alpha=\beta $, the rigidly moving domain-wall solution is exact at arbitrary current densities, leading to an infinite Walker threshold current. The latter becomes finite and decreases with $\beta<\alpha$, approaching a finite value $j_{t0}$ at $\beta=0$ \cite{tataraPRL04}, see Fig.~\ref{num}. In the absence of a strong disorder pinning centers, as assumed so far, $j_{t0}\propto K_\perp$ (which is also the case with the Walker threshold \textit{field} in the absence of an applied current \cite{schryerJAP74}), with an average velocity that slightly above the threshold reads
\begin{equation}
\langle\dot{X}\rangle\propto\sqrt{j^2-j^2_{t0}}\,.
\end{equation}
See the $\beta=0$ curve in Fig.~\ref{num}. At finite $\beta$, the depinning current is determined by the pinning fields, which should be included into the effective field (\ref{H}). The domain-wall velocity at currents slightly above the depinning current is predicted in Ref.~\cite{barnesPRL05} to grow linearly with $j$.

So far in our discussion, we have completely disregarded the random noise contribution to the magnetization dynamics. As noted above, see Eq.~(\ref{hh}), thermal fluctuations are ubiquitous in dissipative systems. Below the (zero-temperature) depinning currents, applied currents can drive the domain wall with finite average velocity $\langle v\rangle$ only by thermal activation. The question how $\ln\langle v\rangle$ scales with the current at low temperatures and weak currents is of fundamental interest beyond the field of magnetism. Experiments on thermally-activated domain-wall motion in magnetic semiconductors \cite{yamanouchiPRL06,yamanouchiSCI07} reveal a ``creep" regime \cite{lemerlePRL98}, in which the effective thermal-activation barrier diverges at low current density $j$, so that $\ln\langle v\rangle$ scales as ${\rm const}-j^{-\mu}$, with an exponent $\mu\sim1/3$. This is inconsistent with the theory based on the Walker ansatz for rigid domain-wall motion \cite{tataraAPL05}, which yields a simple linear scaling of the effective activation barrier and $\ln\langle v\rangle\propto {\rm const}+j$. A refinement of the Walker-ansatz treatment \cite{duinePRL07} cannot explain the experiments either. A scaling theory of creep motion close to the critical temperature \cite{yamanouchiSCI07} does offer a qualitative agreement with measurements by Yamanouchi \textit{et al.} \cite{yamanouchiSCI07}. However, the intrinsic spin-orbit coupling 
in $p$-doped (Ga,Mn)As leads to current-driven effects beyond the standard spin-transfer theories, see, e.g., Refs.~\cite{nguyenPRL06,nguyenPRL07}, which needs to be understood better in the present context.

Even at zero temperature, there are stochastic spin-torque sources in the presence of an applied current, which stem from the discreteness of the angular momentum carried by electron spins, in analogy with the telegraph-like shot noise of electric current carried by discrete particles. A theoretical study of the combined thermal and shot-noise contributions to the stochastic torques for inhomogeneous magnetic configurations~\cite{forosUP07} did not yet explore consequences for the domain-wall dynamics. For example, it is not known whether shot noise assists the current-driven domain-wall depinning at low temperatures. Questions along these lines pose challenging problems for future research.

Effects beyond the theory discussed above are generated by nonadiabatic spin torques, which lead to higher-order in $\partial_t$ and $\boldsymbol{\nabla}$ terms in the equation of motion (\ref{llab}). It is in principle possible to extend linear-response diagrammatic Green's function calculation \cite{tserkovPRB06md,skadsemPRB07,kohnoJPSJ06} by systematically calculating higher-order terms as an expansion in the small parameters, i.e., spin-wave frequency and momentum \cite{thorwartCM07}. A dynamic correction to the spin torque in Eq.~(\ref{llab}) has been found in Refs.~\cite{tserkovPRB06md,thorwartCM07}, which comes down to replacing $\beta\to\beta+n(\hbar/\Delta_{\rm xc})\partial_t$, where $\Delta_{\rm xc}$ is the ferromagnetic exchange splitting and $n=1\,(2)$ for the Stoner ($s-d$) model \cite{tserkovPRB06md,thorwartCM07}. Since this term scales like $\partial_t\boldsymbol{\nabla}$, it is symmetric under time reversal and therefore  nondissipative. Although this dynamic correction is rather small at the typical FMR frequencies $\hbar\omega\ll\Delta_{\rm xc}$, it can cause significant effects at large currents \cite{thorwartCM07}. Starting from an inhomogeneous equilibrium configuration \cite{piechonPRB07,vanhaverbekePRB07,kohnoJPSJ07,thorwartCM07}, such as a magnetic spiral or a domain wall, one can capture nonadiabatic terms in the equation of motion that vanish in linear response with respect to the uniform magnetization considered in Refs.~\cite{tserkovPRB06md,kohnoJPSJ06,skadsemPRB07,duinePRB07}.

For strongly-inhomogeneous magnetic structures, perturbative expansions around a uniform magnetic state fail. For example, for sharp domain walls, the effective equations (\ref{was}) describing wall dynamics and displacement acquire a new term, which can be understood as a force transferred by electrons reflected at the potential barrier caused by the domain wall \cite{bergerJAP84,tataraPRL04}. Electron reflection at a domain wall increases the resistance. Adiabaticity implies a vanishing intrinsic domain-wall resistance (see, however, Ref.~\cite{nguyenPRL06} for a model with strong intrinsic spin-orbit coupling). The force term,  becomes important only for abrupt walls with width $W\sim\lambda_{\rm xc}\equiv\hbar v_F/\Delta_{\rm xc}$. Such nonadiabatic effects are not expected to be strong in metallic ferromagnets, where typically $W\gg\lambda_{\rm xc}\sim\lambda_F$ (the Fermi wavelength). Dilute magnetic semiconductors [such as (Ga,Mn)As] are a different class of materials with longer $\lambda_{\rm xc}$ and a strong spin-orbit coupling \cite{nguyenPRL07}. 

In metallic systems, effects of the spin-torque in the most relevant regime of slow dynamics, $\hbar\omega\ll\Delta_{\rm xc}$, with smooth walls, $W\gg\lambda_{\rm xc}$, and at moderate applied currents  is in our opinion captured by the adiabatic terms linear in $\partial_t$ and $\boldsymbol{\nabla}$. We will now discuss the microscopic basis for Eq.~(\ref{llab}) containing such terms.

\section{Microscopic theory of magnetization dynamics}
\label{mt}

Once the phenomenological equation for current-driven magnetization dynamics is reduced to the form (\ref{llab}), which requires smooth magnetization variation, slow dynamics, and isotropic ferromagnetism, the remaining key questions concern the magnitude and relation between the two dimensionless parameters $\alpha$ and $\beta$. The size of the Gilbert damping constant $\alpha$ is a long-standing open question in solid-state physics, and even a brief review of the relevant ideas and literature is beyond the scope of this paper. A recent model calculation highlighting the multitude of relevant energy scales that control magnetic damping can be found in Ref.~\cite{hankiewiczPRB07}. Here, we discuss only the ratio $\beta/\alpha$, since it is of central importance for macroscopic current-driven phenomena. As noted above, the ratio $\beta/\alpha$ determines, for example, the onset of the ferromagnetic current-driven instability [see Eq.~(\ref{jc})] as well as the Walker threshold current (both diverging when $\beta/\alpha\to1$). The subthreshold current-driven domain-wall velocity is proportional to $\beta/\alpha$ [see Eq.~(\ref{v})], while $\beta/\alpha=1$ is a special point, at which the effect of a uniform current density $\mathbf{j}$ on the magnetization dynamics is eliminated in the frame of reference that moves with velocity $\mathbf{v}=-\mathcal{P}\mathbf{j}/s_0$ [see Eq.~(\ref{gi})]. Although the exact ratio $\beta/\alpha$ is a system-dependent quantity, some qualitative aspects not too sensitive to the microscopic origin of these parameters have recently been discussed  \cite{tserkovPRB06md,kohnoJPSJ06,skadsemPRB07}.

In Ref.~\cite{tserkovPRB06md}, we developed a self-consistent mean-field approach, in which itinerant electrons are described by a time-dependent single-particle Hamiltonian
\begin{equation}
\mathcal{\hat{H}}=\left[ \mathcal{H}_{0}+U(\mathbf{r},t)\right] \hat{1}+\frac{\gamma\hbar}{2}\boldsymbol{\hat{\sigma}}\cdot \left( \mathbf{H}+\mathbf{H}_{\mathrm{xc}}\right)(\mathbf{r},t)+\mathcal{\hat{H}}_\sigma\,,
\label{HKS}
\end{equation}
where the unit matrix $\hat{1}$ and the vector of the Pauli matrices $\boldsymbol{\hat{\sigma}}=(\hat{\sigma}_{x},\hat{\sigma}_{y},\hat{\sigma}_{z})$ form a basis for the Hamiltonian in spin space. $\mathcal{H}_{0}$ is the crystal Hamiltonian including kinetic and potential energy. $U$ is the scalar potential consisting of disorder and applied electric-field contributions. The total magnetic field consists of the applied, $\mathbf{H}$, and exchange, $\mathbf{H}_{\rm xc}$, fields. Finally, the last term in the Hamiltonian, $\mathcal{\hat{H}}_\sigma$, accounts for spin-dephasing processes, e.g, due to quenched magnetic disorder or spin-orbit scattering associated with impurity potentials. This last term is responsible for low-frequency dissipative processes affecting $\alpha$ and $\beta$ in the collective equation of motion (\ref{llab}).

In time-dependent spin-density-functional theory \cite{rungePRL84,capellePRL01,qianPRL02} of itinerant ferromagnetism, the exchange field $\mathbf{H}_{\mathrm{xc}}$ is a functional of the time-dependent spin-density matrix 
\begin{equation}
\rho_{\alpha\beta}(\mathbf{r},t)=\langle\Psi_\beta^\dagger(\mathbf{r})\Psi_\alpha(\mathbf{r})\rangle_t
\end{equation}
that should be computed self-consistently from the Schr\"{o}dinger equation
corresponding to $\mathcal{\hat{H}}$. The spin density of conducting electrons is given by
\begin{equation}
\mathbf{s}(\mathbf{r})=\frac{\hbar}{2}\mbox{Tr}\left[ \boldsymbol{\hat{\sigma}}\hat{\rho}(\mathbf{r})\right]\,.
\label{s}
\end{equation}
Focusing on low-energy magnetic fluctuations that are long range and transverse, we restrict our attention to a single parabolic band. Consideration of realistic band structures is possible from this starting point. We adopt the adiabatic local-density approximation (ALDA, essentially the Stoner model) for the exchange field: 
\begin{equation}
\gamma\hbar \mathbf{H}_{\text{xc}}[\hat{\rho}](\mathbf{r},t)\approx \Delta_{\mathrm{xc}}\mathbf{m}(\mathbf{r},t)\,,
\label{Hxc}
\end{equation}
with direction $\mathbf{m}=-\mathbf{s}/s$ locked to the time-dependent spin density (\ref{s}) (assuming $\gamma>0$).

In another simple model of ferromagnetism, the so-called $s$-$d$ model, conducting $s$ electrons interact with the exchange field  of the $d$ electrons which are assumed to be localized to the crystal lattice sites. The $d$-orbital electron spins are supposed to account for most of the magnetic moment. Because $d$-electron shells have large net spins and strong ferromagnetic correlations, they are usually treated classically. In a mean-field $s$-$d$ description, therefore, conducting $s$ orbitals are described by the same Hamiltonian (\ref{HKS}) with an exchange field (\ref{Hxc}). The differences between the Stoner and $s$-$d$ models for the magnetization dynamics are rather minor and subtle. In the ALDA/Stoner model, the exchange potential is (on the scale of the magnetization dynamics) instantaneously aligned with the total magnetization. In contrast, the direction unit vector $\mathbf{m}$ in the $s$-$d$ model corresponds to the $d$ magnetization, which is allowed to be misaligned with the $s$ magnetization, transferring torque between the $s$ and $d$ magnetic moments. Since most of the magnetization is carried by the latter, the external field $\mathbf{H}$ couples mainly to the $d$ spins, while the $s$ spins respond to and follow the time-dependent exchange field (\ref{Hxc}). As $\Delta_{\rm xc}$ is usually much larger than the external (including demagnetization and anisotropy) fields that drive collective magnetization dynamics, the total magnetic moment will always be very close to $\mathbf{m}$. A more important difference of the philosophy behind the two models is the presumed shielding of the $d$ orbitals from external disorder. The reduced coupling with dissipative degrees of freedom would imply that their dynamics are much less damped. (Whether this is actually the case in real systems remains to be proven, however.) Consequently, the magnetization damping has to come from the disorder experienced by the itinerant $s$ electrons. As in the case of the itinerant ferromagnets, the susceptibility has to be calculated self-consistently with the magnetization dynamics parametrized by $\mathbf{m}$. For more details on this model, we refer to Refs.~\cite{tserkovAPL04,tserkovPRB06md}. With the above differences in mind, the following discussion is applicable to both models. In order to avoid confusion, we remark that the equilibrium spin density $s_0$ introduced earlier refers to the total spin density, i.e., $d$- plus $s$-electron spin density, while Eq.~(\ref{s}) refers only to the latter. The Stoner model is more appropriate for transition-metal ferromagnets because of the strong hybridization between $d$ and $s,p$ electrons. Magnetic semiconductors are characterized by deep magnetic impurity states for which the $s$-$d$ model may be a better choice. 

The single-particle itinerant electron response to electric and magnetic fields in Hamiltonian (\ref{HKS}) is all that is needed to compute the magnetization dynamics microscopically. As mentioned above, the distinction between the Stoner and $s$-$d$ models will appear only at the end of the day, when we self-consistently relate $\mathbf{m}(\mathbf{r},t)$ to the itinerant electron spin response. Before proceeding, we observe that since the constants $\alpha$ and $\beta$ which parametrize the magnetic equation of motion (\ref{llab}) affect the linear response to a small transverse applied field with respect to a uniform magnetization, we can obtain them by a linear-response calculation for the single-domain bulk ferromagnet. The large-scale magnetization texture associated with a domain wall does not affect the value of these parameters, in the considered limit. The linear response to a small magnetic field is complicated by the presence of an electrically-driven applied current, however. Since the Kubo formula based on two-point equilibrium Green's functions is insufficient to calculate the response to simultaneous magnetic and electric fields, we chose to pursue a nonequilibrium (Keldysh) Green's function formalism in Refs.~\cite{tserkovPRB06md,skadsemPRB07}. A technically impressive equilibrium Green's function calculation has been carried out in Ref.~\cite{kohnoJPSJ06}, which to  a large extent confirmed our results, but also contributed some important additions that will be discussed below.

The central quantity in the kinetic equation approach \cite{tserkovPRB06md,skadsemPRB07} is the nonequilibrium component of the $2\times2$ distribution function $\hat{f}_{\mathbf{k}}(\mathbf{r},t)$. In the quasiparticle approximation, valid when $\Delta_{\rm xc}\ll E_F$ \cite{tserkovPRB06md}, the kinetic equation can be reduced to a semiclassical Boltzmann-like equation that accounts for electron drift in response to the electric field as well as the spin precession in the magnetic field. The nonequilibrium component of the spin density reads $\mathbf{s}^\prime=(\hbar /2)\int d^3k\,\mathbf{f}_{\mathbf{k}}/(2\pi)^3$, where $\mathbf{f}_{\mathbf{k}}=\mbox{Tr}[\hat{f}_{\mathbf{k}}\hat{\boldsymbol{\sigma}}]$:
\begin{equation}
\partial_t\mathbf{s}^\prime-\frac{\Delta_{\mathrm{xc}}}{\hbar}\mathbf{z}\times\mathbf{s}^\prime-\frac{\Delta_{\mathrm{xc}}s}{\hbar}\mathbf{z}\times\mathbf{u}=-\frac{\hbar}{2}\int\frac{d^3k}{(2\pi)^3}(\mathbf{v}_{\mathbf{k}}\cdot \partial_{\mathbf{r}})\mathbf{f}_{\mathbf{k}}-\frac{\mathbf{s}^\prime+s\mathbf{u}}{\tau_{\sigma}}\,.
\label{ds}
\end{equation}
$s$ here is the equilibrium spin density of itinerant electrons, $\mathbf{v}_\mathbf{k}=\partial_\mathbf{k}\varepsilon_{\mathbf{k}s}/\hbar$ is the momentum-dependent group velocity, and the magnetization direction $\mathbf{m}=\mathbf{z}+\mathbf{u}$ is assumed to undergo a small precession $\mathbf{u}$ relative to the uniform equilibrium direction $\mathbf{z}$. The first term on the right-hand side is the spin-current divergence and the last term is the spin-dephasing term introduced phenomenologically in Ref.~\cite{tserkovPRB06md} and studied  microscopically in Ref.~\cite{skadsemPRB07}. As detailed in Ref.~\cite{tserkovPRB06md}, the spin currents have to be calculated from the full kinetic equation and then inserted in Eq.~(\ref{ds}). The final result (for the Stoner model) is given by Eq.~(\ref{lla}) or, equivalently, Eq.~(\ref{llab}), with $\alpha=\beta$. The latter is proportional to the spin-dephasing rate:
\begin{equation}
\beta=\frac{\hbar}{\tau_\sigma \Delta_{\rm xc}}\,.
\end{equation}
The derivation assumes $\omega,\tau^{-1}_\sigma\ll\Delta_{\rm xc}/\hbar$, which is typically the case in real materials sufficiently below the Curie temperature. The $s$-$d$ model yields the same result for $\beta$, but
\begin{equation}
\alpha=\eta\beta
\label{ab}
\end{equation}
is reduced by the $\eta=s/s_0$ ratio, i.e., the fraction of the itinerant to the total angular momentum. [Note that Eq.~(\ref{ab}) is also valid for the Stoner model since then $s_0=s$.] For the $s$-$d$ model, the equation of motion (\ref{llab}) clearly cannot be reduced to Eq.~(\ref{lla}), since $\alpha\neq\beta$. The steady-state current-driven velocity (\ref{v}) for both mean-field models becomes
\begin{equation}
v=-\frac{\beta\mathcal{P}j}{\alpha s_0}=-\frac{\mathcal{P}j}{s}\,,
\label{vel}
\end{equation}
where $s$ is the itinerant electron spin density. Interestingly, the velocity (\ref{vel}) is completely determined by properties of the conducting electrons, even for the $s-d$ model. In the Drude model,
\begin{equation}
v\propto\frac{E\tau}{m^\ast}\,,
\label{vdrift}
\end{equation}
where $E$ is the applied electric field, $\tau$ is the characteristic momentum scattering time, and $m^\ast$ is the effective mass of the itinerant bands at the Fermi energy. We expect the velocity (\ref{vdrift}), which is essentially the conducting electron drift velocity, to be suppressed for the $s$-$d$ model if the $d$ orbitals are coupled to their own dissipative bath, which has not been included in the above treatment.

Ref.~\cite{kohnoJPSJ06} refines these results by relaxing the assumption that $\Delta_{\rm xc}\ll E_F$ and by considering also anisotropic spin-dephasing impurities, which results in $\alpha\neq\beta$ for both Stoner and $s$-$d$ models. Ref.~\cite{duinePRB07} later offered a Keldysh functional-integral approach leading to the same results. (These authors also found stochastic torques expressed in terms of thermal fluctuations (\ref{hh}) in the weak current limit; see, however, Ref.~\cite{forosUP07} for additional current-induced stochastic terms present in the case of an inhomogeneous magnetization.) Consider, for example, weak magnetic disorder described by the potential $\hat{H}_\sigma=\mathbf{h}(\mathbf{r})\cdot\hat{\boldsymbol{\sigma}}$ with Gaussian white-noise correlations $\langle h_a(\mathbf{r})h_b(\mathbf{r}^\prime)\rangle\propto U_a\delta_{ab}\delta(\mathbf{r}-\mathbf{r}^\prime)$, where $U_a=U_\perp$  ($U_\parallel$) when $\mathbf{a}$ is perpendicular (parallel) to the equilibrium magnetization direction. (Spin-orbit interaction associated with scalar disorder gives similar results.) For isotropic disorder, $U_\perp=U_\parallel$, and $\Delta_{\rm xc}\ll E_F$$, \alpha/\beta\approx\eta$ with $\eta=s/s_0$, as was already discussed (reducing to $\eta=1$ for the Stoner model). Even for larger exchange, the correction to this $\alpha/\beta$ ratio turns out to be rather small: For parabolic bands, for example, $\alpha/\beta\approx[1-(\Delta_{\rm xc}/E_F)^2/48]\eta$. This ratio is more sensitive to anisotropies $U_\parallel\neq U_\perp$, however, so that in general $\alpha/\beta\neq\eta$ even in the limit $\Delta_{\rm xc}/E_F\to0$ \cite{kohnoJPSJ06}.

\section{Summary and outlook}

Our microscopic understanding is based on a mean-field approximation, in which itinerant electrons interact self-consistently with a space- and time-dependent exchange field. We presented results for the local-spin-density approximation and the mean-field $s$-$d$ model. We identified a relation between dissipative terms parametrized by $\alpha$ and $\beta$ and spin-dephasing scattering potentials. The central result for the collective low-frequency long-wavelength current-driven magnetization dynamics can be formulated as a generalization of the phenomenological Landau-Lifshitz-Gilbert equation, accounting for the current-driven torques. One should in general also include stochastic terms due to thermal fluctuations as well as nonequilibrium shot-noise contribution in the presence of applied current $\mathbf{j}$ \cite{forosUP07}. Despite some recent efforts, stochastic effects remain to be relatively unexplored both theoretically and experimentally, however.

The most important parameter that determines the effect of an electric current on the collective magnetization dynamics in extended systems is the ratio $\beta/\alpha$. We find that this ratio is not universal and in general depends on details of the band structure and spin-dephasing processes. Nevertheless, simple models give $\alpha\sim\beta$ with the special limit $\alpha\approx\beta$ for the Stoner model with weak and isotropic spin-dephasing disorder. Solving the magnetization equation of motion for a domain wall is rather straightforward at low dc currents, when the wall is only slightly compressed. The domain-wall motion can then be modeled within the Walker ansatz, based on parametrizing the magnetic dynamics in terms of  wall position and spin distortion. Two regimes can then be distinguished: At the lowest currents, the wall moves steadily in the presence of a constant uniform current, while above the so-called Walker threshold, the magnetization close to the wall center starts oscillating, resulting in a singular dependence of the average velocity on the applied current.

The values of the $\alpha$ and $\beta$ parameters are not affected by the magnetization textures. Micromagnetic simulations can provide better understanding of experimental results in the regimes where domain walls are not well described by a one-dimensional model. Experiments can contribute to the understanding by studying ferromagnets with systematic variations of impurity types and concentration, for Py and other different materials. Experimental investigation of creep in metallic ferromagnets at temperatures far below the critical ones, as compared to studies \cite{yamanouchiPRL06,yamanouchiSCI07} on magnetic semiconductors close to the Curie transition, are highly desirable in order to advance our understanding.

Besides realistic microscopic evaluations of the key parameters $\alpha$ and $\beta$, the collective current-driven magnetization dynamics pose many theoretical challenges, in the spirit of classical nonlinear dynamical systems. Current-driven magnetism displays a rich behavior well beyond what can be achieved by applied magnetic fields only. At supercritical currents, ferromagnetism becomes unstable, possibly leading to chaotic dynamics \cite{liJAP05}, although alternative scenarios have been also suggested \cite{shibataPRL05}. Domain-wall dynamics in a medium with disordered pinning potentials pose an interesting yet, at weak applied currents, tractable problem. Spin torques and dynamics in sharp walls and the role of strong intrinsic spin-orbit coupling (relevant for dilute magnetic semiconductors) are not yet completely understood. Oscillatory domain-wall motion under ac currents and in curved geometries is also starting to attract attention both experimentally and theoretically~\cite{thomasNAT06,krugerPRB07}. Another direction of recent activities concern the backaction of a moving domain wall on the charge degrees of freedom \cite{barnesPRL07,duineCM07dw,stamenovaCM07,yangCM07,tserkovCM07em}.

With the exciting recent and forthcoming experimental developments, the questions concerning interactions of the collective ferromagnetic order with electric currents will certainly challenge theoreticians for many years to come. The prospects of using purely electric means to efficiently manipulate magnetic dynamics are also promising for practical applications.

\section{Acknowledgments}

We would like to thank the Editors for carefully reading the manuscript and making many useful comments. Discussions with Hans-Joakim Skadsem, Hiroshi Kohno, and Sadamichi Maekawa are gratefully acknowledged. This work has been supported by EC Contract IST-033749 \textquotedblleft DynaMax.\textquotedblright

\end{document}